\documentclass[sigconf]{acmart}
\usepackage[nolist]{acronym}
\usepackage{graphicx}
\usepackage{subcaption}
\usepackage{wrapfig}
\usepackage{hyperref}
\AtBeginDocument{%
  \providecommand\BibTeX{{%
    \normalfont B\kern-0.5em{\scshape i\kern-0.25em b}\kern-0.8em\TeX}}}

\copyrightyear{2022} 
\acmYear{2022} 
\setcopyright{acmlicensed} 
\acmConference[UIST '22 Adjunct]{The Adjunct Publication of the 35th Annual ACM Symposium on User Interface Software and Technology}{October 29-November 2, 2022}{Bend, OR, USA}
\acmBooktitle{The Adjunct Publication of the 35th Annual ACM Symposium on User Interface Software and Technology (UIST '22 Adjunct), October 29-November 2, 2022, Bend, OR, USA}
\acmDOI{10.1145/3526114.3558739}
\acmISBN{978-1-4503-9321-8/22/10}

\begin{document}

\begin{acronym}[]
	\acro{AI}{Artificial Intelligence}
	\acro{ACC}{Adaptive Cruise Control}
	\acro{ADAS}{Advanced Driver Assistance System}
	\acro{CAN}{Controller Area Network}
	\acro{ECU}{Electronic Control Unit}
	\acro{GDPR}{General Data Protection Regulation}
	\acro{GOMS}{Goals, Operators, Methods, Selection rules}
	\acro{HCI}{Human-Computer Interaction}
	\acro{HMI}{Human-Machine Interaction}
	\acro{HU}{Head Unit}
	\acro{IVIS}{In-Vehicle Information System}
	\acro{KLM}{Keystroke-Level Model}
	\acro{KPI}{Key Performance Indicator}
	\acro{LKA}{Lane Keeping Assist}
	\acro{LCA}{Lane Centering Assist}
	\acro{MGD}{Mean Glance Duration}
	\acro{OEM}{Original Equipment Manufacturer}
	\acro{OTA}{Over-The-Air}
	\acro{AOI}{Area of Interest}
	\acro{RF}{Random Forest}
	\acro{SA}{Steering Assist}
	\acro{SHAP}{SHapley Additive exPlanation}
	\acro{TGD}{Total Glance Duration}
	\acro{UCD}{User-centered Design}
	\acro{UX}{User Experience}
\end{acronym}

\title[ICEBOAT]{ICEBOAT: An Interactive User Behavior Analysis Tool for Automotive User Interfaces}




\author{Patrick Ebel}
\email{ebel@cs.uni-koeln.de}
\orcid{0000-0002-4437-2821}
\affiliation{%
	\institution{University of Cologne}
	\city{Cologne}
	\country{Germany}}

\author{Kim Julian Gülle}
\email{k.guelle@campus.tu-berlin.de}
\orcid{}
\affiliation{%
	\institution{TU Berlin}
	\city{Berlin}
	\country{Germany}}

\author{Christoph Lingenfelder}
\email{christoph.lingenfelder@mercedes-benz.com}
\orcid{0000-0001-9417-5116}
\affiliation{%
	\institution{MBition GmbH}
	\city{Berlin}
	\country{Germany}}

\author{Andreas Vogelsang}
\email{vogelsang@cs.uni-koeln.de}
\orcid{0000-0003-1041-0815}
\affiliation{%
	\institution{University of Cologne}
	\city{Cologne}
	\country{Germany}}

\renewcommand{\shortauthors}{Ebel et al.}

\begin{abstract}
In this work, we present ICEBOAT an interactive tool that enables automotive UX experts to explore how users interact with \acp{IVIS}. Based on large naturalistic driving data continuously collected from production line vehicles, ICEBOAT visualizes drivers' interactions and driving behavior on different levels of detail. Hence, it allows to easily compare different user flows based on performance- and safety-related metrics.
\end{abstract}



\maketitle


\section{Introduction}

In modern vehicles, large center stack touchscreens have become an integral part of the driver-vehicle interaction. Today's \acp{IVIS} offer a variety of functions that are increasingly similar to those of tablets and smartphones. Therefore, \acp{IVIS} are now subject to the same high demands when it comes to usability.

To keep pace with this rapid development and to continuously develop interfaces that are well received by the customers, there is an increased need for data-driven support in the automotive UX design~\cite{ebel.2021, Ebel.2020a, Orlovska.2018}. Whereas websites and apps track every interaction a user makes and design decisions are made in consideration of conversion rates, time on task, or error rates \cite{King.2017}, the decision-making process for the design of \acp{IVIS} is more complex. During manual and partially automated driving, the driver must monitor the driving situation at all times \cite{SAE.2021}, making driving the primary task. Hence, the interaction with \acp{IVIS} is only the secondary task. While interacting with IVISs drivers need to distribute their attention between the primary driving task and the secondary touchscreen task. This behavior is correlated with an increased crash risk~\cite{dingus.2016}. Thus, the metrics used to compare different designs or the basis on which to decide which features to prioritize are substantially different from what we are used to in web or app development~\cite{Harvey.2016}. Automotive UX experts have to consider the interdependencies between driving behavior, interaction behavior, and glance behavior~\cite{Klauer.2006}.

In previous research~\cite{ebel.2021, Ebel.2020a}, we identified needs and potentials to support the automotive design process by providing feature usage statistics, user flow analyses, and context-dependent visualizations. Based on these findings, we developed three standalone hierarchical user behavior visualizations that proved their usefulness in a user study~\cite{ebel.2021a}. However, to leverage the potential of such visualization they need to be integrated into the product design process. 

Therefore, this work in progress presents ICEBOAT, an \textbf{I}nter- a\textbf{C}tive Us\textbf{E}r \textbf{B}ehavi\textbf{O}r \textbf{A}nalysis \textbf{T}ool that enables UX experts to analyze in-vehicle user flows and driving-related performance metrics interactively based on continuously collected data from production line vehicles. The contribution of this work is twofold: First, we conducted semi-structured interviews to gather insights from practitioners on how large-scale field user interaction data needs to be integrated into their design process. Second, using a co-design approach we build an interactive web app that allows UX experts to analyze drivers' touchscreen interactions on different levels of granularity.


\begin{figure*}
	\centering
	\includegraphics[width = \linewidth]{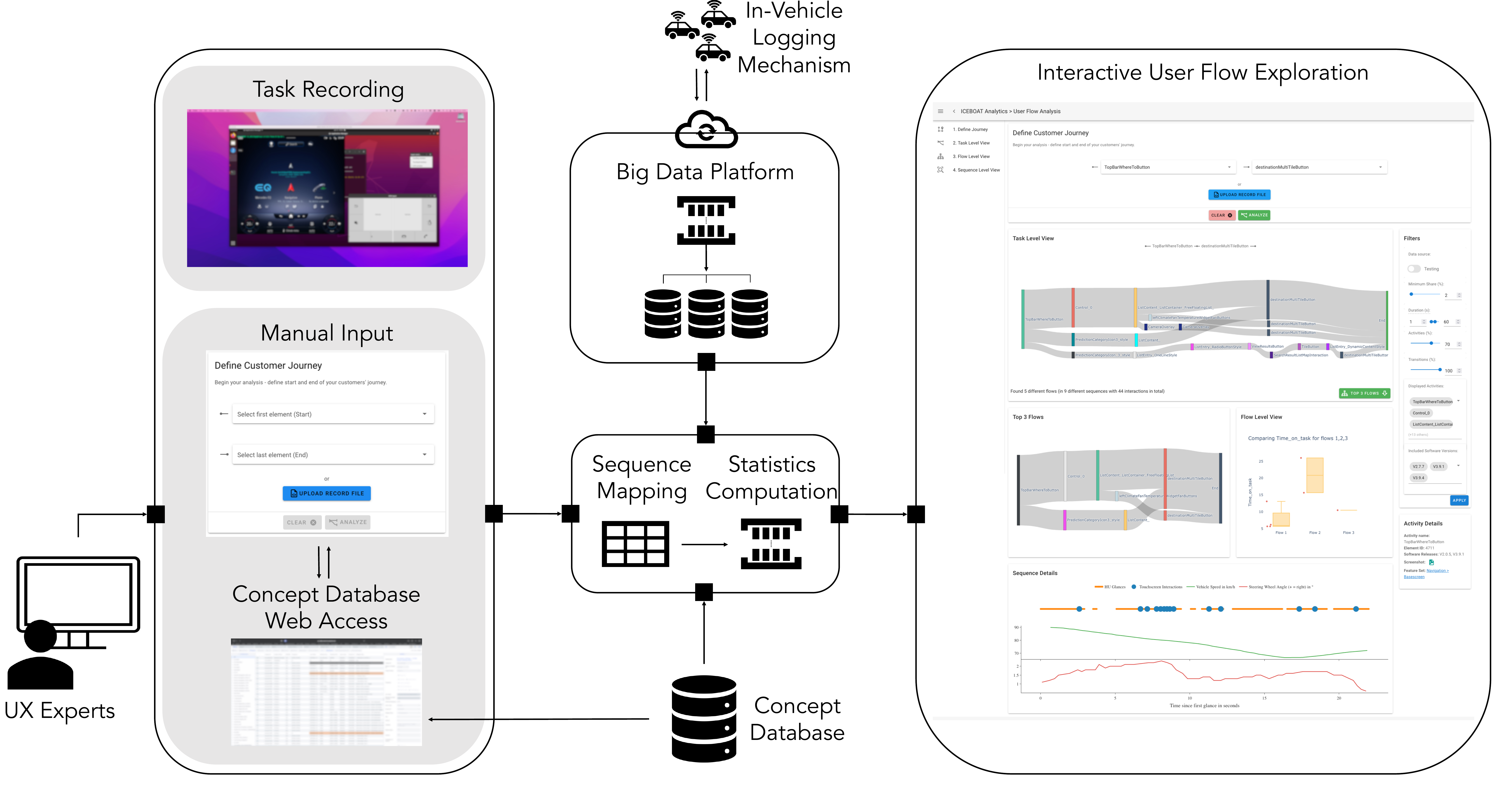} 
	\caption{The ICEBOAT system architecture. } 
	\label{fig:Architecture} 
\end{figure*}

\section{Methodology}
We develop ICEBOAT by following a user-centric design approach. First, we conducted 4 interviews to elicit initial requirements. The participants had at least 5 years of relevant professional experience and have been with our research partner for more than 5 years. Therefore, we consider them knowledgeable regarding the automotive-specific design processes. After an initial introduction, we presented the visualizations introduced by \citeauthor{ebel.2021a}~\cite{ebel.2021a}. We then asked the interviewees how they would improve these visualizations and how they envision an interactive tool that would benefit their daily work. After eliciting the requirement, we designed a first prototype of the ICEBOAT tool. We improve this prototype continuously based on the user feedback we get in co-design sessions.

\section{ICEBOAT}

ICEBOAT is an interactive tool that enables automotive UX experts to explore how users interact with \ac{IVIS}. Based on large naturalistic driving data collected continuously from production line vehicles, it provides various options to analyze drivers' interaction and driving behavior. The system architecture is shown in Figure~\ref{fig:Architecture}. 

ICEBOAT's UI provides users with two different options to start their analysis: They can either choose (1) \textit{Manual Input} or (2) \textit{Task Recording}. When using the manual input option, users are asked to define the task they want to analyze by selecting the first and last UI element out of a searchable drop-down list. 
To ease this process, users can also access the concept database using a separate web interface and search for the respective UI elements. The concept database contains all interactive UI elements of the IVIS software and further information like the application (e.g. navigation, climate, etc.) they belong to, the screen on which they are placed, or the function that they trigger. However, as there are thousands of different UI elements, the users are not necessarily aware of all respective identifiers.
Therefore, we developed the task recording input option. It enables users with limited knowledge to playfully define the task they are interested in. Users can click through an emulated version of the IVIS, similar to what is displayed to drivers on the center stack touchscreen in the car. Once they find a task that they want to analyze, they can record the corresponding interactions. The emulation environment then outputs a file containing the identifiers of the respective UI elements. As of now, this file can then be uploaded to ICEBOAT which uses the first and last interaction of the recorded sequence as the start and end of a task. 

After the users define the task they want to analyze, ICEBOAT extracts all relevant interaction sequences from the customer data. The data used for the analysis is collected \ac{OTA} from all production line vehicles of our research partner that are equipped with the most recent software architecture. The \textit{In-vehicle Logging Mechanism} collects touchscreen interactions from the UI interface and driving-related data from the vehicle bus. This information is continuously sent to the \textit{Big Data Platform} where it is further processed such that it serves the needs of ICEBOAT. After sequence extraction, the data is enhanced with additional information from the concept database and user flow statistics are calculated. Finally, the results are computed and visualized according to the three levels introduced by \citeauthor{ebel.2021a}~\cite{ebel.2021a}: the \textit{Task Level View}, \textit{Flow Level View}, and \textit{Sequence Level View}. 

In the course of the initial study, we found that users use the Task Level View as a reference to which they come back after they went into the other views for more detailed analyses \cite{ebel.2021a}. We, therefore, choose this view as the entry point of the visual exploration journey. The visualization hierarchy then unfolds dynamically as users dive deeper into more specific analyses. In the Task Level View, the interaction data is aggregated and visualized in form of an adapted Sankey diagram, allowing users to easily assess which interactions drivers performed to solve the chosen task. Users can also choose between various filtering options allowing them to, for example, only show data for the most prominent user flows, specific software versions, or car models.
If users are further interested in how different flows that belong to the same task compare, they can choose the respective flows and define a metric of their choice used for comparison. The Flow Level View then shows the distribution of interaction sequences according to the chosen metric in form of boxplots. The underlying datapoints are visualized next to the boxplot. Each dot represents one interaction sequence. If users are interested in a specific sequence, (e.g. an outlier with a very long interaction time) and the driving situation in which the interactions took place, they can click on the visualization and are re-directed to the Sequence Level View. This view shows an overlay of the respective touchscreen interactions and driver glances. Furthermore, driving-related signals like the vehicle speed and steering wheel angle are visualized. This allows users to better judge the specific situation as in-vehicle interactions are highly context-sensitive and need to be evaluated based on the driving situation.

\section{Conclusion and Future Work}
In this work in progress, we introduce ICEBOAT, an interactive user behavior analysis tool for automotive user interfaces. ICEBOAT processes telematics data that is continuously collected from production line vehicles. The tool enables UX experts not only to evaluate how drivers interact with the in-vehicle touchscreen but also allows them to analyze user interactions in light of the driving situation. The proposed implementation is based on previous research \cite{ebel.2021, ebel.2021a} and represents the current result of an ongoing co-design process. In future iterations, we aim to introduce new visualizations, implement functionalities that allow users to save analyses, and introduce more metrics and filters. The final tool will be evaluated in an extensive usability study.

\section{Acknowledgements}
We want to thank Fabian Ober for his work on the user flow recording option.

\balance
\bibliographystyle{ACM-Reference-Format}
\bibliography{iceboat.bib}

\end{document}